\documentclass[12t]{article}

\usepackage[cp1251]{inputenc}

\usepackage{amsmath}

\begin{document}


\newcommand{\N}{N\raise.7ex\hbox{\underline{$\circ $}}$\;$}

\renewcommand{\theequation}{\thesection.\theequation}
\numberwithin{equation}{section}

\thispagestyle{empty}

{\em International Conference on

Non-Euclidean Geometry and its  applications.

5 -- 9 July 2010, Kluj-Napoca (Kolozv\'{a}r), ROMANIA}

\vspace{2mm}

\begin{center}

{\bf  E.M. Ovsiyuk, V.M. Red'kov. \\[1mm]
EXACT SPECTRUM FOR QUANTUM OSCILLATOR IN SPACES OF CONSTANT CURVATURE FROM  WKB-QUANTIZATION\\[1mm]
Institute of Physics \\
National Academy of Sciences of Belarus
}

\end{center}

\begin{quotation}

Quantum-mechanical WKB-method is elaborated for the known quantum oscillator problem
in curved 3-spaces models Euclid, Rie\-mann, and Lobachevsky    $E_{3}, H_{3}, S_{3}$ in
the framework of the complex variable  function theory.
Generalized generally covariant Schr\"{o}dinger equation is considered.
In all three space models, exact energy levels are found with the help of constructing special formal WKB-sieries.

\end{quotation}

\section{ Introduction}

It is well known that energy spectrum of the hydrogen atom had been calculated long begore creating the comprehensive
quantum mechanical theory: Bohr
\cite{1913-Bohr, 1914-Bohr, 1915-Bohr}, Sommerfeld
\cite{1915-Sommerfeld(1), 1915-Sommerfeld(2), 1916-Sommerfeld,
1919-Sommerfeld}, Wilson  {1915-Wilson(1), 1922-Wilson(2)}, Ishiwara
 \cite{1915-Ishiwara},  Planck  \cite{1915-Planck,
1916-Planck}, Schwarzschild \cite{1916-Schwarzschild}, Epstein
\cite{1916-Epstein(1)},  Wentzel \cite{1926-Wentzel}, Brillouin
\cite{1922-Brillouin, 1926-Brillouin(1)}. It was established that the Bohr-Sommerfeld rules,
 basis of the "old" \hspace{1mm} quantum mechanics even without rigorous mathematical foundation,
 are closely related  to the so-called WKB-approximation in the consistent quantum theory: see in
Langer \cite{1937-Langer}, Titchmarsh  \cite{1954-Titchmarsh},
Ponomarev \cite{1966-Ponomarev}.

 Looking for exactly solvable  models in the framework of "new" \hspace{1mm} quantum theory,
 some coolness towards approximate
  (all the more without foundation) methods  and any achievements  of the Bohr-Sommerfeld mechanics  was inevitable.
But  the same question arises in the literature:  why in the case of hydrogen atom the Bohr-Sommerfeld rule leads to
 the known  exact energy spectrum. Also,  from time to time in the literature one can face
  the statement of the sort: in a potential
 $\varphi$ the Bohr-Sommerfeld quantization gives an exact result $\epsilon_{n}(\varphi)$:
Bailey  \cite{1964-Bailey},
Froman and Froman \cite{1965-Froman-Froman}, Krieger \cite{1966-Krieger},
Rosenzweig and Krieger  \cite{1968-Rosenzweig-Krieger}, Nisio
\cite{1973-Nisio}, Elutin and Krivchenkov \cite{1974-Elutin-Krivchenkov},
Voros  \cite{1976-Voros,
1983-Voros,1983-Voros, 2000-Voros}, De
Witt and Morette \cite{1976-De Witt-Morette}, Neveu  \cite{1976-Neveu}, Gomes et al
\cite{1986-Gomes-Thomaz-Vasconcelos}, Dutt et al  \cite{1986-Dutt-Khare-Sukhatme},
Lemos and Natividade \cite{1987-Lemos-Natividade},  Schopf \cite{1988-Schopf},  Katayama
\cite{1992-Katayama}, Kobylinsky er al
\cite{1990-Kobylinsky-Stepanov-Tutik}, Fujii and Funahashi
\cite{1995-Fujii-Funahashi}, Robnik and Salasnich
\cite{1997-Robnik-Salasnich(1),1997-Robnik-Salasnich(2)}, Delabaere et al
\cite{1997-Delabaere-Dillinger-Pham}, Kudryashov and Vanne  \cite{2001-Kudryashov-Vanne}.

In the present work we turn  to an oscillator problem but now  placed concur\-rently in three different
curved space backgrounds: Euclid $E_{3}$ (zero curvature),  Lobachevsky (negative constant curvature) $H_{3}$, and
Riemann $S_{3}$ (positive cons\-tant curvature).
We have considered Schr\'{o}dinger's equation.
 It is shown that there can be constructed special  WKB-series that provide us with exact spectra
 by taking into account only two firs terms of these series, in all three models  $E_{3}, H_{3}, S_{3}$.
This work continues two earlier considerations \cite{1986-Otchik-Red'kov} and  \cite{2003-Red'kov}
of the analogous problem for hydrogen atom in  space models $E_{3}, H_{3}, S_{3}$,
 motivation and general mathematical
techniques are similar.

\section{ Oscillator in $E_{3}$ and WKB-quantization }

Let us  consider a non-relativistic oscillator in Euclid  space model $E_{3}$.
In Schr\"{o}din\-ger's equation, the variables are separated by the known substitution
 $ \Psi (r , \theta, \phi ) = f(r ) \;
Y_{lm}(\theta, \phi)$:
\begin{eqnarray}
 {d ^{2} f \over d r^{2}} + {2 \over  r } {d f \over d r}  + \left [ {2M
\over \hbar^{2}} \;  ( E - {1 \over 2} k \;  r^{2}
 ) - {l(l+1) \over r^{2}
} \; \right  ] \; f (r ) = 0  \; . \label{1.1}
\end{eqnarray}

\noindent Let $t$ be a new variable: $ t: \;  r^{2}  = e^{2t}$, eq.
 (\ref{1.1})  takes the form
 \begin{eqnarray}
 ({d^{2} \over dt^{2} } + {d \over dt })\;R
+ \left [ \; {2M  \over \hbar^{2}} \;  ( E e^{2t}  - {1 \over 2} k
e^{4t}  ) - l(l+1) \; \right   ] \; f (r ) = 0  \; . \nonumber
\end{eqnarray}

\noindent Excluding the  first derivative term by a substitution
 $R = e^{-t/2}\; S (t)$
\begin{eqnarray}
 {d^{2} \over dt^{2} } S + \;
\left [ \; {   2M E  e^{2t}  - M k  \; e^{4t}   - \hbar^{2}l(l+1)
\over \hbar^{2}  }
  -{1 \over 4} \; \right ] \; S = 0  \; .
\label{1.2}
\end{eqnarray}

\noindent With the notation
\begin{eqnarray}
- \hbar^{2} l(l+1) =  - \hbar^{2} \; [ (l+1/2)^{2} - {1 \over 4} ]
= - L^{2} + { \hbar^{2}  \over 4} \; ,\;\; A = -Mk  \; , \nonumber
\\
 B = 2M E  \; , \;\;    C = - L^{2} \; , \;\;
\Pi^{2}(t) =   A\;  e^{4t} + B \;  e^{2t}   +C
 \nonumber
 \label{1.4a}
 \end{eqnarray}

\noindent from eq. (\ref{1.2}) we get
\begin{eqnarray}
{d^{2} \over dt^{2}}\; S(t)  + \;  {\Pi^{2}(t) \over \hbar^{2}}
 \;S (t) = 0 \; .
\label{1.4b}
\end{eqnarray}

\noindent Now let us expand  the function $Q(t)$ into a series in terms of $(\hbar / i)^{n}$
\begin{eqnarray}
S(t) = \mbox{exp} \; [ \; {i\over \hbar } \; \int \; Q(t) dt \;  ]
\; , \nonumber
\\
{\hbar \over i} \; {d \over dt} Q + Q^{2} - \Pi^{2}  = 0\; , \;\;
Q(t) = \sum_{n=0}^{\infty} \;  [ \; ({\hbar \over i})^{n} \;
Q_{n}(t)\; ] \; , \nonumber
\\
Q_{0} = \sqrt{\Pi^{2}} \; , \; Q_{1} = -{1 \over 2Q_{0}} \; Q'_{0}
\; , \;\; Q_{2} = -{1 \over 2Q_{0}}\; ( \;Q'_{1} + Q_{1}^{2}  \; )
\; , \nonumber
\\
Q_{n} = -\; {1 \over 2 Q_{0}} \; ( \; {d \over dt} Q_{n-1} \; + \;
\sum_{k=1}^{n-1}  \; Q_{n-k}  Q_{k} \;  ) = 0 \; , \; n = 3,4,5,
... ;
\label{1.5}
\end{eqnarray}

\noindent the symbol  $^{\prime}$  denotes   derivative $d /dt$.

We  will assume that  the wave   $S(r) $ corresponding to a bound state, being considered as a  function of
complex variable $t$, has a finite number of of zeros in the complex plane,  which  are allocated  at real axis
between  classical turning pints.
According  the known theorem in complex  variable function theory  the  number of such zeros of $S(t)$ within  certain domain
 can be calculated  through derivative
 $(\ln S(t))^{\prime}$ along a contour bounding that domain
\begin{eqnarray}
{1 \over 2\pi i} \; \oint_{{\cal L}} [{d \over d t} \ln S(t) ] \;
dt = n \;  ,
\nonumber
\end{eqnarray}
\noindent from whence substituting  a series instead of $Q(t)$, we  arrive at
\begin{eqnarray}
\sum_{n=0}^{\infty} \; [ \;
({\hbar \over i})^{n} \; \oint _{\cal L} \; Q_{n}(t) \;dt \;   ]
\; = 2 \pi \hbar \; n \; . \label{1.6a}
\end{eqnarray}

\noindent It should be especially emphasized that relationship (\ref{1.6a}) is a precise  mathema\-tical  condition without
any approximation. Accounting for only two first terms leads to the Bohr-Sommerfeld
quantization rule
\begin{eqnarray}
\oint _{\cal L} \;  Q_{0}(t) dt + {\hbar \over i}  \oint _{\cal L}
\;  \; Q_{0}(t) dt \approx   2 \pi \hbar \; n \; .
\label{1.6b}
\end{eqnarray}

\noindent Calculation of the integrals is reduced to finding  residues
in two points
\begin{eqnarray}
\oint _{\cal L} \; { Q_{n}(z) \over z} \;dz \;  = ( - 2 \pi i )\;
\sum \; \mbox{res}\; _{z = 0,  \infty} \;  { Q_{n}(z) \over z}  \;
.
\label{1.6c}
\end{eqnarray}

\noindent The contribution of the first order term is
\begin{eqnarray}
\oint _{{\cal L}} \; Q_{0}(t) \; d t =  2 \pi   \; (\; -i \sqrt{C}
-i\; {B \over 2  \sqrt{A} } \; )\; .
\label{1.7}
\end{eqnarray}

\noindent The contribution of the second  order term is
\begin{eqnarray}
{\hbar \over i}\;   \; \oint _{{\cal L}} Q_{1}(t) dt  = {\hbar
\over i}\;  (- {1 \over 2}) \; (-2 \pi i) \;  \times
\nonumber
\\
\times
\sum \; \mbox{res}\;
_{z = 0,  \infty} \; {1 \over 2}\;  {4 A z^{4} + 2B z^{2}  \over z
( A z^{4}   + B z^{2} + C ) }  =  -2 \pi \hbar \; . \nonumber
\label{1.8}
\end{eqnarray}

\noindent Therefore, the  quantization rule   (\ref{1.6b}
)  gives
\begin{eqnarray}
2\pi \;  (  - i \sqrt{C} - i\; { B \over 2 \sqrt{A} } - \hbar
 ) = 2\pi \; (2n) \; ,
 \nonumber
 \end{eqnarray}

\noindent from whence it follows the exact  energy spectrum
\begin{eqnarray}
E =  \hbar\; \sqrt{{ k \over M }} \; (2n + l + 3/2) \;.
\label{1.9}
\end{eqnarray}

\section{Oscillator in hyperbolic   model $H_{3}$ }

In Lobachevsky  space, the Scr\"{o}dinger equation for an oscillator problem
\begin{eqnarray}
 ( -\; {\hbar^{2} \over 2M} \; \Delta_{2} \;  +
{1 \over 2} k \rho^{2} \mbox{th}^{2}\;  r \; ) \; \Psi = E \; \Psi
\; \nonumber
\end{eqnarray}

\noindent  after separation of the variables
$ \Psi (r ,
\theta, \phi ) = f(r ) \; Y_{lm}(\theta, \phi)$ leads to
\begin{eqnarray}
 {d ^{2} f \over d r^{2}} + {2 \over \mbox{th}\; r } {d  f \over d r} +
\left [ {2M\rho^{2} \over \hbar^{2}}   ( E - {1 \over 2} k
\rho^{2} \mbox{th}^{2}  r
 ) - {l(l+1) \over \mbox{sh}^{2} r
}  \right ]  f  = 0 \;   . \label{2.1}
\end{eqnarray}

It should ne noted special symmetry property  of the radial equation
with respect to the change $r \rightarrow -r$. Negative values for $r$ are non-physical, but
below in taking into account zeros of complex variable function we must take into consideration
 these non-physical zeros as well.

 Let  $t$ be a new variable: $ \mbox{th}^{2}\; r = e^{2t}$,  eq.  (\ref{2.1}) takes the form
 \begin{eqnarray}
 ({d^{2} \over dt^{2} } + {d \over dt })f
+ {1 \over (1 - e^{2t})^{2} }  \left [  {2M\rho^{2} \over
\hbar^{2}}   ( E e^{2t}  - {1 \over 2} k \rho^{2} e^{4t}  ) -
l(l+1)  ( 1 - e^{2t} )  \right   ]  f = 0 \;   . \nonumber
\end{eqnarray}

\noindent Excluding the first derivative term by a substitution
 $f = e^{-t/2}\; S (t)$
\begin{eqnarray}
 {d^{2} \over dt^{2} } S + \;
\left [ \; {   2M E \rho^{2} \;  e^{2t}  - M k \rho^{4}   \;
e^{4t}   - \hbar^{2}l(l+1) \; ( 1 - e^{2t} ) \over \hbar^{2}\; (1
- e^{2t})^{2}  }
  -{1 \over 4} \; \right ] \; S = 0  \; .
\label{2.2}
\end{eqnarray}

\noindent Using the notation
\begin{eqnarray}
- \hbar^{2} l(l+1) =  - \hbar^{2} \; [ (l+1/2)^{2} - {1 \over 4} ]
= - L^{2} + { \hbar^{2}  \over 4} \; , \nonumber
\\
A = -Mk \rho^{4} \; , \qquad B = 2M E \rho^{2} - \hbar^{2} +L^{2}
\; , \qquad   C = - L^{2} \; , \nonumber
\\
\Pi^{2}(t) = {  A\;  e^{4t} + B \;  e^{2t}   +C \over (1 -
e^{2t})^{2}  }
 \; , \qquad
\Delta (t) =  {5 - e^{2t}  \over 4 }  { e^{2t} \over (1 -
e^{2t})^{2} } \; ,
\label{2.3}
\end{eqnarray}

\noindent we arrive at
\begin{eqnarray}
{d^{2} \over dt^{2}}\; S(t)  + \; [\;  {\Pi^{2}(t) \over
\hbar^{2}} \;+ \; \Delta (t)   \; ] \;S (t) = 0 \; . \label{2.4}
\end{eqnarray}

\noindent Further we follow the standard procedure
\begin{eqnarray}
S(t) = \mbox{exp} \; [ \; {i\over \hbar } \; \int \; Q(t) dt \;  ]
\; , \nonumber
\\
{\hbar \over i} \; {d \over dt} Q + Q^{2} - \Pi^{2} + ( {\hbar
\over i}) ^{2} \Delta = 0\; , \;\; Q(t) = \sum_{n=0}^{\infty} \; [
\; ({\hbar \over i})^{n} \; Q_{n}(t)\; ] \; , \nonumber
\\
Q_{0} = \sqrt{\Pi^{2}} \; , \; Q_{1} = -{1 \over 2Q_{0}} \; Q'_{0}
\; , \;\; Q_{2} = -{1 \over 2Q_{0}}\; [ \;Q'_{1} + Q_{1}^{2} +
\Delta \; ] \; , \nonumber
\\
Q_{n} = -\; {1 \over 2 Q_{0}} \; [ \; {d \over dt} Q_{n-1} \; + \;
\sum_{k=1}^{n-1}  \; Q_{n-k}  Q_{k} \;  ] = 0 \; , \; n = 3,4,5,
... \label{2.5}
\end{eqnarray}

\noindent The quantization condition (exact  that)
take the form
\begin{eqnarray}
{1 \over 2\pi i} \; \oint_{{\cal L}} [{d \over d t} \ln S(t) ] \;
dt = 2n \;  \
\nonumber
\end{eqnarray}
\noindent or
\begin{eqnarray}
\;\; \sum_{n=0}^{\infty} \; [ \; ({\hbar \over i})^{n} \; \oint
_{\cal L} \; Q_{n}(t) \;dt \; ] \; = 2 \pi \hbar \; (2n ) \; .
\label{2.6a}
\end{eqnarray}

\noindent In particular, allowing for only two first terms of the WKB-series gives
\begin{eqnarray}
\oint _{\cal L} \;  Q_{0}(t) dt + {\hbar \over i}  \oint _{\cal L}
\;  \; Q_{0}(t) dt \approx   2 \pi \hbar \; (2n) \; . \label{2.6b}
\end{eqnarray}

In calculating  the contour integrals one is to use the variable
 $z
= e^{t}  = \mbox{th}\; r$,  correspondingly the contour ${\cal L}(z)$
including zeros of the function can be presented  by the Figure
\begin{center}
Fig.  1  (Integration contour  ${\cal L}(z)$)
\end{center}

\vspace{-7mm} \unitlength=0.6mm
\begin{picture}(160,40)(-80,0)
\special{em:linewidth 0.4pt} \linethickness{0.4pt}

\vspace{-5mm} \put(-80,0){\vector(+1,0){160}}  \put(+75,-8){$
\mbox{th}\; r $} \put(0,-30){\vector(0,+1){60}}

\put(+5,-5){\line(+1,0){50}} \put(+5,+5){\line(0,-1){8}}
\put(+55,+5){\line(-1,0){50}} \put(+55,-5){\line(0,+1){10}}

\put(-5,-3){\line(+1,0){10}}

\put(-5,+5){\line(-1,0){50}} \put(-55,-5){\line(+1,0){60}}
\put(-55,+5){\line(0,-1){10}} \put(-5,+5){\line(0,-1){8}}

\put(-60,0){\circle*{2}}  \put(-65,-7){$-1$}
\put(+60,0){\circle*{2}}  \put(+58,-7){$+1$}

\end{picture}
\vspace{18mm}

\noindent and we are to  find residues in four points
\begin{eqnarray}
\oint _{\cal L} \; { Q_{n}(z) \over z} \;dz \;  = ( - 2 \pi i )\;
\sum \; \mbox{res}\; _{z = 0, \pm 1, \infty} \;  { Q_{n}(z) \over
z}  \; . \label{2.6b}
\end{eqnarray}

The contribution of the first order term is
\begin{eqnarray}
\oint _{{\cal L}} \; Q_{0}(t) \; d t =  2 \pi   \; (\; -i \sqrt{C}
+i  \sqrt{A+B+C} +i \sqrt{A} \; )\; . \nonumber \label{2.7}
\end{eqnarray}

\noindent
The contribution of the second  order term is
\begin{eqnarray}
{\hbar \over i}\;   \; \oint _{{\cal L}} Q_{1}(t) dt  =
  {\hbar \over i}\;  (- {1 \over 2}) \; (-2 \pi i) \; \times
  \nonumber
  \\
  \times
  \sum \;
\mbox{res}\; _{z = 0, \pm 1, \infty} \; \left [\; {1 \over 2}\; {4
A z^{4} + 2B z^{2}  \over z ( A z^{4}   + B z^{2} + C ) } + {2
z^{2} \over z( 1 -z^{2} ) }  \right  ] \; ; \nonumber
\end{eqnarray}

\noindent therefore we get
\begin{eqnarray}
{\hbar \over i}\;   \; \oint _{{\cal L}} Q_{1}(t) dt  = {\hbar
\over i} \;  \; (- {1 \over 2} )  (-2 \pi i) \; ( -2 )  =  -2 \pi
\hbar \; . \nonumber
\end{eqnarray}

\noindent Thus, the Bohr-Sommerfeld rule gives
\begin{eqnarray}
 - \sqrt{-C} +   \sqrt{-A-B-C} + \sqrt{-A}   -  \hbar \approx
\hbar \;(2 n)  \; , \label{2.9a}
\end{eqnarray}

\noindent from whence it follows
\begin{eqnarray}
\sqrt{ Mk\rho^{4} - 2M E \rho^{2} + \hbar^{2}} = \hbar(2n + l
+{3\over 2})   - \sqrt{Mk \rho^{4} } \; .\nonumber
\end{eqnarray}

However, we know an exact quantization condition (from exact solution of the differential equation
(\ref{2.1})
in hypergeometric functions -- see for example \cite{1992-Katayama})
\begin{eqnarray}
 \sqrt{ Mk\rho^{4} - 2M E \rho^{2} + \hbar^{2}}
=  \hbar \; (2n + l +{3\over 2} )  - \sqrt{Mk \rho^{4}  +
{\hbar^{2} \over 4}}  \; , \label{2.10b}
\end{eqnarray}

\noindent below it will be more convenient to use dimensionless form
\begin{eqnarray}
= + \sqrt{ \mu - 2 \epsilon  + 1} = 2n + l + {3 \over 2 }  -  {
\sqrt{1 + 4 \mu } \over 2}  \; .
 \label{2.10c}
\end{eqnarray}

It is the point  of  primary importance. Turning back to starting equation  (\ref{2.2}),  one may note that
from the very beginning in this equation a special formal rearrangement  should  be performed
\begin{eqnarray}
 {d^{2} \over dt^{2} } S  -{1 \over 4} \; S + { 1 \over \hbar^{2} (1 - e^{2t})^{2}  }
   \left [   \; ( 2M E \rho^{2}  + \hbar^{2} \beta - \hbar^{2} \beta) +
e^{2t}  \right.  \nonumber
\\
\left. +
 ( - M k \rho^{4} + \hbar^{2} \alpha - \hbar^{2} \alpha )   e^{4t}   - \hbar^{2}l(l+1) ( 1 - e^{2t} )
 \; \right ]  \;  S = 0  \; ,
\nonumber
\\
A = -Mk\rho^{4} +  \hbar^{2} \;  \alpha \; , \qquad B = 2ME\rho^{2} +
\hbar^{2} \; \beta  +  \L^{2} \;, \nonumber \\
 C = - L^{2} \; ,
\qquad  \alpha = -  {1 \over  4 } \; , \qquad  \beta =  {5 \over 4} \; ,
\qquad
 \alpha + \beta = 1 \; , \nonumber \label{2.11}
\end{eqnarray}

\noindent which should give different representation  of the main  differential equation
\begin{eqnarray}
 {d^{2} \over dt^{2} } S + \;
\left [  {  ( 2M E \rho^{2}  + \hbar^{2} \beta )   e^{2t} +
 ( - M k \rho^{4} + \hbar^{2} \alpha )   e^{4t}   - \hbar^{2}l(l+1) ( 1 - e^{2t} )
\over \hbar^{2} (1 - e^{2t})^{2}  } \right. \nonumber
\\
\left.
  -{1 \over 4}  - {5 \over 4}  {e^{2t} \over (1 - e^{2t})^{2} }
    + {1 \over 4}  {e^{4t} \over (1- e^{2t})^{2}} \;
   \right ]  S = 0  \; .
\nonumber
\\
\label{2.12}
\end{eqnarray}

Correspondingly, we have expressions for $A,B,C$ and
 $\Delta (t)$
\begin{eqnarray}
 \Pi^{2}(t) = {  A\;  e^{4t} + B \;  e^{2t}   +C \over (1 -
e^{2t})^{2}  } , \;\;  \Delta =  -{1 \over 4}  - {5 \over 4} \;
{e^{2t} \over (1 - e^{2t})^{2} }
    + {1 \over 4} \; {e^{4t} \over (1- e^{2t})^{2}} \;   ,
\nonumber
\\
A = -Mk \rho^{4} - {1 \over 4} \hbar^{2}   \; , \qquad B = 2M E
\rho^{2} - {5 \over 4}\hbar^{2}  + L^{2} \; , \qquad   C = - L^{2}
\; ; \nonumber
\\
\label{2.13}
\end{eqnarray}

\noindent  from whence the Bohr-Sommerfeld rule results the exact
energy spectrum (let   $l+ {3\over 2} + 2n =N$)
\begin{eqnarray}
2 \epsilon  = - N^{2}  + \sqrt{1 + 4 \mu } \; N     + {3 \over 4}  \; ,
\label{2.14}
\end{eqnarray}

\noindent
or differently
\begin{eqnarray}
2 \epsilon  -1 = + \mu - (N - {\sqrt{1 + 4 \mu }\over 2})^{2} \;
\; .
\label{2.15}
\end{eqnarray}

\section{Oscillator in spherical space $S_{3}$ }

In Riemann spherical space , the Scr\"{o}dinger equation for an oscillator problem
\begin{eqnarray}
 ( -\; {\hbar^{2} \over 2M} \;
\Delta_{2} \;  + {1 \over 2} k \rho^{2} \mbox{tg}^{2}\;  r \; ) \;
\Psi = E \; \Psi \; \nonumber
\end{eqnarray}

\noindent  after separation of the variables
 $ \Psi (r ,
\theta, \phi ) = f(r )  Y_{lm}(\theta, \phi)$ gives
\begin{eqnarray}
{d ^{2} f \over d r^{2}} + {2 \over \mbox{tg}\; r } {d f \over d
r} + \left [ {2M\rho^{2} \over \hbar^{2}}   ( E - {1 \over 2} k
\rho^{2}  \mbox{tg}^{2}\;  r
 ) - {l(l+1) \over \mbox{sin}^{2} r
}  \right  ]   f  = 0  \; . \label{3.1}
\end{eqnarray}

\noindent
Again, take notice symmetry   $r \rightarrow
-r$.
 Let  $r$ be a new variable:
$ \mbox{tg}^{2}\; r = e^{2t}$,
 eq.   (\ref{3.1}) takes the form
(let if be $f = e^{-t/2} S (t)$ ):
\begin{eqnarray}
 {d^{2} \over dt^{2} } S + \;
\left [  {   2M E \rho^{2}   e^{2t}  - M k \rho^{4}   \;
e^{4t}   - \hbar^{2}l(l+1)  ( 1 + e^{2t} ) \over \hbar^{2} (1
+ e^{2t})^{2}  }
  -{1 \over 4}  \right ]  S = 0  \; .
\label{3.2}
\end{eqnarray}

\noindent With  the notation
\begin{eqnarray}
- \hbar^{2} l(l+1) =  - \hbar^{2} \; [ (l+1/2)^{2} - {1 \over 4} ]
= - L^{2} + { \hbar^{2}  \over 4} \; , \nonumber
\\
A = -Mk \rho^{4} \; , \qquad B = 2M E \rho^{2} + \hbar^{2} -L^{2}
\; , \qquad   C = - L^{2} \; , \nonumber
\\
\Pi^{2}(t) = {  A\;  e^{4t} + B \;  e^{2t}   +C \over (1 +
e^{2t})^{2}  }
 \; , \qquad
\Delta (t) =  -{5 + e^{2t}  \over 4 }  { e^{2t} \over (1 +
e^{2t})^{2} }
 \label{3.4a}
\end{eqnarray}

\noindent eq. (\ref{3.2}) reads
\begin{eqnarray}
{d^{2} \over dt^{2}}\; S(t)  + \; [\;  {\Pi^{2}(t) \over
\hbar^{2}} \;+ \; \Delta (t)   \; ] \;S (t) = 0 \; . \label{3.4b}
\end{eqnarray}

\noindent Further we follow the above procedure
\begin{eqnarray}
{1 \over 2\pi i} \; \oint_{{\cal L}} [{d \over d t} \ln S(t) ] \;
dt = 2n \;  \;
\nonumber
\end{eqnarray}
or \begin{eqnarray}
 \sum_{n=0}^{\infty} \; [ \;
({\hbar \over i})^{n} \; \oint _{\cal L} \; Q_{n}(t) \;dt \;   ]
\; = 2 \pi \hbar \; (2n ) \; . \label{3.6a}
\end{eqnarray}

\noindent The Bohr-Sommerfeld rule reads
\begin{eqnarray}
\oint _{\cal L} \;  Q_{0}(t) dt + {\hbar \over i}  \oint _{\cal L}
\;  \; Q_{0}(t) dt \approx   2 \pi \hbar \; (2n) \; . \label{3.6b}
\end{eqnarray}

\noindent
Calculating the contour integrals is  reduced to calculating residues in four points in complex plane
( $z
= e^{t}  = \mbox{tg}\; r$)
\begin{eqnarray}
\oint _{\cal L} \; { Q_{n}(z) \over z} \;dz \;  = ( - 2 \pi i )\;
\sum \; \mbox{res}\; _{z = 0, \pm i, \infty} \;  { Q_{n}(z) \over
z}  \; . \label{3.6c}
\end{eqnarray}

\noindent
The Bohr-Sommerfeld rule  leads to
\begin{eqnarray}
 - \sqrt{-C} +   \sqrt{-A+B-C} - \sqrt{-A}   -  \hbar \approx
\hbar \;(2 n)  \; .
 \nonumber \label{3.9}
\end{eqnarray}

\noindent  and further
we arrive at the relationship
\begin{eqnarray}
\sqrt{ Mk\rho^{4} + 2M E \rho^{2} + \hbar^{2}} = \hbar (2n + l
+{3\over 2} ) + \sqrt{Mk \rho^{4} } \; . \label{3.10a}
\end{eqnarray}

However, we know an exact quantization condition (from exact solution of the differential equation
(\ref{2.1})
in hypergeometric functions)
\begin{eqnarray}
 \sqrt{ Mk\rho^{4} + 2M E \rho^{2} + \hbar^{2}}
=  \hbar \; (2n + l +{3\over 2} )  + \sqrt{Mk \rho^{4}  +
{\hbar^{2} \over 4}} \label{3.10b}
\end{eqnarray}

\noindent or in dimensionless notation
\begin{eqnarray}
 \sqrt{ \mu + 2 \epsilon  + 1} = 2n + l + {3 \over 2 }  +  {
\sqrt{1 + 4 \mu } \over 2}  \; . \label{3.10c}
\end{eqnarray}

Turning back to starting equation  (\ref{3.2}),  one may note that
from the very beginning in this equation a special formal rearrangement  should  be performed
\begin{eqnarray}
 {d^{2} \over dt^{2} } S  - {1 \over 4}\; S + {1 \over  \hbar^{2} (1 + e^{2t})^{2}  }\;
 \left [\;
   ( 2M E \rho^{2}  + \hbar^{2} \beta - \hbar^{2} \beta)
e^{2t} \right.
\nonumber
\\
\left.  +
 ( - M k \rho^{4} + \hbar^{2} \alpha - \hbar^{2} \alpha )   e^{4t}   - \hbar^{2}l(l+1) ( 1 + e^{2t} )\right ] \;
    S = 0   ,
\nonumber \end{eqnarray}
\begin{eqnarray}
\ A = -Mk\rho^{4} - \hbar^{2} \; \alpha \; , \qquad B =
2ME\rho^{2} +   \hbar^{2} \; \beta  - \L^{2} \;, \nonumber
\\
 C = - L^{2} \; , \;\;
\alpha =   -{1 \over  4 } \; , \; \beta =  {5 \over 4} \; , \;\;
\alpha + \beta = 1 \;, \nonumber
\end{eqnarray}

\noindent
which should give different representation  of differential equation
\begin{eqnarray}
 {d^{2} \over dt^{2} } S + \;
\left [  {  ( 2M E \rho^{2}  +\hbar^{2} \beta )   e^{2t} +
 ( - M k \rho^{4} - \hbar^{2} \alpha )   e^{4t}   - \hbar^{2}l(l+1) ( 1 + e^{2t} )
\over \hbar^{2} (1 + e^{2t})^{2}  } \right. \nonumber
\\
\left.
  -{1 \over 4}  - {5 \over 4} \; {e^{2t} \over (1 + e^{2t})^{2} }
   - {1 \over 4} \; {e^{4t} \over (1+ e^{2t})^{2}} \;
   \right ]  S = 0  \; .
\nonumber
\\
\label{3.12}
\end{eqnarray}

\noindent so that  $A,B,C$ and  $\Delta (t)$ are
\begin{eqnarray}
\Pi^{2}(t) = {  A\;  e^{4t} + B \;  e^{2t}   +C \over (1 -
e^{2t})^{2}  } \;, \;\;    \Delta =  -{1 \over 4}  - {5 \over 4}
\; {e^{2t} \over (1 + e^{2t})^{2} }
    - {1 \over 4} \; {e^{4t} \over (1+ e^{2t})^{2}} \;   .
\nonumber
\\
A = -Mk \rho^{4} + {1 \over 4} \hbar^{2}   \; , \qquad B = 2M E
\rho^{2} + {5 \over 4}\hbar^{2}  -L^{2} \; , \qquad   C = - L^{2}
\; ; \nonumber
\\
\label{3.13}
\end{eqnarray}

\noindent
from whence the Bohr-Sommerfeld rule results the exact
energy spectrum (let   $l+ {3\over 2} + 2n =N$)
\begin{eqnarray}
 2\epsilon  = (l +  2n +  {3\over 2}  )^{2}  + \sqrt{1 +4 \mu} \;  (l +  2n +  {3\over 2}  )
  - {3 \over 4} \; .
\label{3.14}
\end{eqnarray}

\noindent
or differently
\begin{eqnarray}
2 \epsilon  +1 =  \mu + (N + {\sqrt{1 + 4 \mu }\over 2})^{2} \;  .
\label{3.15}
\end{eqnarray}

Let summarize result:

It is shown that in the spaces of Lobachevsky and Riemann, similar
to the case  od Euclidean model  $E_{3}$, in   WKB-theory for
harmonic oscillator potential there  can be constructed special
WKB-series, such that only two first terms give a non-zero
contribution into  the Bohr -- Sommerfeld quantization condition
providing us with an exact energy spectrums in all three  models
  $E_{3}, H_{3}, S_{3}$.

\section{Acknowledgement }

The work was supported by the grand  of BRFFI (Belarusian Republican Foundation for
Fundamental Research), No F09K--123.

We wish to thank the Organizers of the  International Conference "Non-Euclidean Geometry and its  applications."
 (5 -- 9 July 2010, Kluj-Napoca (Kolozv\'{a}r), ROMANIA)   for
having given us the opportunity to talk on this subject, as well as HCAA-ESF
 (Harmonic and Complex Analysis and its Applications -- European Science Foundation Research Networking Programme)
 for partial support.


\begin{thebibliography}{xxx}




\bibitem{1913-Bohr}
N. Bohr. On the Constitution of atoms and molecules
// Phil. Mag. 1913. Vol. 26. P. 1 -- 25, 476 -- 502,  857 -- 875


\vspace{-2mm}
\bibitem{1914-Bohr}
N. Bohr. Effet des champs\'{e}lectrique et magn\`{e}tique sur les
raies spectrales // Phil. Mag. 1914. Vol. 27. P. 506.


\vspace{-2.2mm}
\bibitem{1915-Bohr}
N. Bohr. Sur les s\'{e}ries spectrales de l'hydrog\`{e}ne et la
structure de l 'atome.
// Phil. Mag. 1915. Vol. 29. P. 332.


\vspace{-2.2mm}
\bibitem{1915-Sommerfeld(1)}
A. Sommerfeld. Zur Theorie der Balmerschen Serie
// M\"{u}nchener  Berichte. 1915.  S. 425 -- 458.


\vspace{-2.2mm}
\bibitem{1915-Sommerfeld(2)}
A. Sommerfeld. Die Feinstructur der wasserstoff und
wasserstoff\"{a}hnlichen Linien // M\"{u}nchener  Berichte. 1915.
S. 459 -- 500.


\vspace{-2.2mm}
\bibitem{1916-Sommerfeld}
A. Sommerfeld. Zur Quantentheorie der Spektrallinien
//  Annalen der Physik. 1916. Bd. 51. S. 1 -- 94, S. 125 -- 167.



\vspace{-2.2mm}
\bibitem{1919-Sommerfeld}
A. Sommerfeld. Atombau und Spektrallinien. Braunchweig, Vieweg,
1919, S. 327 -- 357; 520 -- 522; Atomic structure and spectral
lines. L.: Methuen, 1923, P. 467 -- 496; 608 -- 611.


\vspace{-2.2mm}
\bibitem{1915-Wilson(1)}
W. Wilson. The quantum theory of radiation and line spectra
// Phil.   Mag. 1915. Vol. 29. P. 795 -- 802.


\vspace{-2.2mm}
\bibitem{1922-Wilson(2)}
W. Wilson. The quantum theory and electromagnetic phenomena
// Proc. Roy. Soc. London. A.  1922.  Vol. 102.  P. 478 -- 483.


\vspace{-2.2mm}
\bibitem{1915-Ishiwara}
J. Ishiwara. Die universelle Bedeutung des Wirkungsquantums.
// Tokio Sugaku Buturigakkawi Kizi. 1915. Bd 8. S. 106 -- 116.



\vspace{-2.2mm}
\bibitem{1915-Planck}
M. Planck. Die Quantenhypothese f\"{u}r  Molekeln mit  mehreren
Freiheitsgraden // Verhandlungen  der Deutschen  Physikalischen
Gesellschaft. 1915. Bd. 17. S. 407 -- 418; S. 438 -- 451.


\vspace{-2.2mm}
\bibitem{1916-Planck}
 M. Planck.  Die  physikalische Struktur  des Phasenraum
// Annalen der Physik. 1916. Bd. 50.  S. 385 -- 418.



\vspace{-2.2mm}
\bibitem{1916-Schwarzschild}
K. Schwarzschild. Zur Quantenhypothese
// Berliner Berichte. 1916.  S. 548 -- 568.


\vspace{-2.2mm}
\bibitem{1916-Epstein(1)}
 P.S. Epstein.
 Zur Theorie des Starkeffektes
// Phys.  Zeit. 1916.  Bd. 17. S. 148 -- 150.







\bibitem{1926-Wentzel}
G. Wentzel. Eine Verallgemeinerung der Quantembedingungen f\"{u}r
die Zwecke der Wellenmechanik // Zeit. Phys.  1926.  Bd. 38. S.
518 -- 529.



\vspace{-2.2mm}
\bibitem{1922-Brillouin}
L. Brillouin. Th\'eorie des quanta et l'atome de Bohr.
 Press Universitaire de France. 1922.



\vspace{-2.2mm}
\bibitem{1926-Brillouin(1)}
L. Brillouin.
 La m\'{e}chanique ondulatoire  de Schr\"{o}dinger; une  m\'{e}thode g\'{e}n\'{e}rale
de r\'{e}solution par approximations successives
// Coptes Rendus Acad. Sci. Paris.  1926. T.  183. P. 24-44.



\vspace{-2.2mm}
\bibitem{1937-Langer}
 R.E. Langer.
 On the connection formulas and the solutions of the wave equation
// Phys. Rev. 1937. Vol. 51. P. 669 -- 676.


\vspace{-2.2mm}
\bibitem{1954-Titchmarsh}
 E.C. Titchmarsh.  On the asymptotic distridution of eigenvalues
 // Quart. J. Math. 1954. Vol. 5. P. 228 -- 240.


\vspace{-2.2mm}
\bibitem{1966-Ponomarev}
L.I. Ponomarev. Lectures in quasiclassics. Kiev, 1966 (in Russian).







\vspace{-2.2mm}
\bibitem{1964-Bailey}
 P.B. Bailey.
 Exact quantization rules for the one-dimensional
Scr\"odinger equation with turning points
// J. Math. Phys. 1964. Vol. 5, N  9. P. 1293 -- 1297.


\vspace{-2.2mm}
\bibitem{1965-Froman-Froman}
N. Froman,  P.O. Froman.
 JWKB Approximation: contributions to the theory.
North-Holland Publ. Co. Amsterdam. 1965.


\vspace{-2.2mm}
\bibitem{1966-Krieger}
J.B. Krieger. Asymptotic properties of perturbation theory
// J. Math. Phys. 1966. Vol. 9,  N  3. P. 432 -- 435.



\vspace{-2.2mm}
\bibitem{1968-Rosenzweig-Krieger}
 C.  Rosenzweig,   J.B. Krieger.
 Exact quantization conditions
// J. Math. Phys. 1968. Vol. 9,  N 6. P. 849 -- 860.


\vspace{-2.2mm}
\bibitem{1973-Nisio}
Sigeko Nisio. The formation of the Sommerfeld quantum theory of
1916
// Jap. Stud. Hist. Sci. 1973,  N 12. P. 54 -- 60.






\vspace{-2.2mm}
\bibitem{1974-Elutin-Krivchenkov}
 P.V. Elutin, D.V. Krivchenkov.
 On applicabilitu os the quasiclasiccal approximation.
// TMF. 1974. Vol. 19. \N 2. P. 233 -- 236 (in Russian).

\vspace{-2.2mm}
\bibitem{1976-Voros}
A.  Voros.
 Semi-classical approximations
// Ann. Inst. H. Poincar\'{e}. A.  1976, Vol. 24. P. 31 -- 90.



\vspace{-2.2mm}
\bibitem{1983-Voros}
A. Voros. The return of the quadratic oscillator. The complex WKB
method
//  Ann. Inst. H. Poincar\'e.  A. 1983.  Vol. 39. P.  211 -- 338.






\vspace{-2.2mm}
\bibitem{1994-Voros}
A. Voros.
 Exact quantization condition for anharmonic oscillators
(in one dimension)
//  J. Phys. A.  1994. Vol. 27. P. 4653 -- 4661.


\vspace{-2.2mm}
\bibitem{2000-Voros}
A. Voros.
 Exercises in exact quantization
 // J. Phys. A. 2000. Vol. 33. P. 7423 -- 7450.




\vspace{-2.2mm}
\bibitem{1976-De Witt-Morette}
De Witt-Morette C.
 The semiclassical expansion.
// Ann. Phys. N.Y. 1976. Vol. 97.  P. 367 -- 399.


\vspace{-2.2mm}
\bibitem{1976-Neveu}
A. Neveu. Semiclassical quantization  methods in field theory
// Phys. Rep. C. 1976. Vol. 23. P. 265 -- 272.






\vspace{-2.2mm}
\bibitem{1986-Gomes-Thomaz-Vasconcelos}
 M.A.F. Gomes,   M.T. Thomaz,   G.L. Vasconcelos.
 Matrix formulation for the Wentzel-Kramers-Brillouin quantization rule
// Phys. Rev. A. 1986. Vol. 34,  N  5. P. 3598 -- 3604.



\vspace{-2.2mm}
\bibitem{1986-Dutt-Khare-Sukhatme}
R. Dutt, A. Khare,  U.P. Sukhatme. Exactness of supersymmetric WKB
spectra  for  shape-invariante potentials
// Phys. Lett. B. 1986.    Vol. 181. P. 295 -- 298.





\vspace{-2.2mm}
\bibitem{1987-Lemos-Natividade}
 N.A. Lemos,   C.P. Natividade.
 Harmonic oscilator in  expanding universes
// Nuovo Cim. B. 1987. Vol. 99,  N  2. P. 211 -- 225.


\vspace{-2.2mm}
\bibitem{1988-Schopf}
H.G. Sch\"{o}pf. Zur Geschichte der Bohr-Sommerfeldschen
Quantentheorie
// Ann. Phys. 1988, Bd. 45, N 8. S. 595 -- 604.


\vspace{-2.2mm}
\bibitem{1992-Katayama}
N. Katayama.
 A note on a  quantum-mechanical  harmonic oscillator in a a space of constant curvature
// Nuovo Cim. B. 1992.  Vol. 107,  N 7. P. 763 -- 768.


\vspace{-2.2mm}
\bibitem{1990-Kobylinsky-Stepanov-Tutik}
 N.A. Kobylinsky,  S.S. Stepanov,   R.S. Tutik.
 Semiclassical approach to ground state within the Klein-Gordon equation
// J. Phys. A. 1990. Vol. 23,  N  6. P. 237 -- 241.


\vspace{-2.2mm}
\bibitem{1995-Fujii-Funahashi}
K.  Fujii, K. Funahashi.
 Ecactness in the WKB-approximation  for some homogeneous spaces
// J. Math. Phys.  1995. Vol. 36. P. 4590 -- 4611;
hep-th/9501145.

\vspace{-2.2mm}
\bibitem{1997-Robnik-Salasnich(1)}
M. Robnik, L. Salasnich.
 WKB exactness for the angular momentum and the Kepler problem: from
the torus quantization to the exact one
// J. Phys. A. 1997.  Vol. 30.  P. 1719 -- 1729; quant-ph/9603014.

\vspace{-2.2mm}
\bibitem{1997-Robnik-Salasnich(2)}
M. Robnik, L. Salasnich. WKB to all orders and the accuracy of the
semiclassical quantization
// J. Phys. A. 1997. Vol. 30. P. 1711 -- 1718; quant-ph/9610027.

\vspace{-2.2mm}
\bibitem{1997-Delabaere-Dillinger-Pham}
E. Delabaere, H. Dillinger and F. Pham. Exact semiclassical
expansions for one-dimensional quantum oscillators
//  J. Math. Phys.  1997.  Vol. 38. P. 6126 -- 6184.


\vspace{-2.2mm}
\bibitem{2001-Kudryashov-Vanne}
V.V. Kudryashov, Yu.V. Vanne.
 Explicit summation of the
constituent  WKB series and new approximate wave functions
// J. Appl. Math.  2002. Vol. 2. P. 265 (2002); arXiv:quant-ph/0102111.




\vspace{-2.2mm}

\bibitem{1986-Otchik-Red'kov}
V.S. Otchik V.M. Red'kov.
    Quantum-mechanical Kepler problem in spve of constant curvarure.
    Preprint 298, IF AN BSSR. Minsk,  1986. 49 pages (in  Russian).

\vspace{-2.2mm}
\bibitem{2003-Red'kov}
 V.M. Red'kov.
    On WKB-quantization in Lobachevski and Riemann 3-spaces.
    // Nonlinear phenomena in  complex systems. 2003, Vol. 6, \N 2. P. 654-668.


\end{thebibliography}
\end{document}